\documentclass[12pt,showpacs,preprintnumbers,amsmath,amssymb]{revtex4}

\usepackage{graphicx}

\begin{document}

\title{Nonequilibrium phase transitions and tricriticality in a three-dimensional lattice system with random-field competing kinetics}

\author{Nuno Crokidakis}
\email{nuno@if.uff.br}
\affiliation{
Instituto de F\'{\i}sica - Universidade Federal Fluminense \\
Av. Litor\^anea s/n \\
24210-340 \hspace{5mm} Niter\'oi - Rio de Janeiro \hspace{5mm} Brazil}

\date{\today}

\begin{abstract}

We study a nonequilibrium Ising model that stochastically evolves under the simultaneous operation of several spin-flip mechanisms. In other words, the local magnetic fields change sign randomly with time due to competing kinetics. This dynamics models a fast and random diffusion of disorder that takes place in dilute metallic alloys when magnetic ions diffuse. We performe Monte Carlo simulations on cubic lattices up to $L=60$. The system exhibits ferromagnetic and paramagnetic steady states. Our results predict first-order transitions at low temperatures and large disorder strengths, which correspond to the existence of a nonequilibrium tricritical point at finite temperature.  By means of standard finite-size scaling equations, we estimate the critical exponents in the low-field region, for which our simulations uphold continuous phase transitions.

\end{abstract}


\maketitle

\section{Introduction}

The Random Field Ising Model (RFIM) is one of the most studied systems in magnetism (for reviews, see \cite{binder_review,belanger_review} and more recently \cite{dotsenko}), because of its mathematical simplicity and because the possibility of reproducting frustration, a phenomenon that occurs in real magnetic systems. In addition, the identification of the RFIM with some diluted antiferromagnets in the presence of a uniform magnetic field \cite{fishman,cardy}, like ${\rm Fe_{x}Zn_{1-x}F_{2}}$ and ${\rm Fe_{x}Mg_{1-x}Cl_{2}}$ \cite{belanger_review,belanger,birgeneau}, have attracted the attention of theoretical and experimental researchers.

The equilibrium RFIM has extensively been studied by different approaches, such as mean-field theory \cite{aharony,schneider,meu_jpcm}, the renormalization group \cite{berker,aizenman} and Monte Carlo (MC) simulations \cite{hernandez,fytas}. However, the critical behavior of the nonequilibrium case \cite{garrido_prl} is little understood, with the majority of the results concerning one-dimensional systems \cite{lacomba1,lacomba2,garrido_epl}. A mean field theory was also developed \cite{alonso}, but numerical results are rare; to the best of our knowledge, only a recent work considered the problem via MC simulations \cite{meu_jstat} $^{1}$.\footnotetext[1]{The nonequilibrium Ising spin glass was studied via MC simulations in \cite{miranda}.}

On the fact that there is a consensus that \textit{equilibrium} quenched models are a suitable representation of some realistic situations in physics, some controversy on basic issues persist, and the interpretation of laboratory experiments by models with quenched disorder is not satisfactory \cite{binder_review,miranda,parisi_book,dickman}. Concerning spin glasses, for example, quenched models like the Edwards-Anderson (EA) \cite{ea} neglect the diffusion of magnetically active ions. Diffusion constantly modifies the distance between each specific pair of spin ions in certain substances like dilute metallic alloys (CuMn, for example) and, consequently, one should probably allow for variations both in space and time of the exchange interactions in a model \cite{miranda}. In the same way, we can imagine other disordered systems in which the random variables change in space and time, like random-field models. These effects do not seem to be correctly described by another class of equilibrium models, namely, the \textit{annealed} systems. For example, the change with time of the spatial distribution of couplings in the annealed version of the EA model \cite{thorpe} is constrained by the need to reach equilibrium with the other degrees of freedom. Therefore, impurities tend to be strongly correlated, which is not observed in most substances \cite{dickman}. As a consequence, the annealed version of the EA model do not exhibits a spin glass state \cite{thorpe}. In addition, the equilibrium random-field models also present problems to describe theoretically experimental results. While annealed and quenched versions of the RFIM predict continuous phase transitions between the ordered and disordered phases  \cite{fytas,hernandez,claudete} $^{2}$\footnotetext[2]{There are some controversies about the order of the low-temperature phase transition in the RFIM, but recent simulations suggest the occurrence of continuous transitions \cite{fytas,hernandez}.}, measurements on diluted antiferromagnets like $\rm{Fe_{x}Mg_{1-x}Cl_{2}}$, which are prototypes of experimental realizations of systems under random fields \cite{belanger_review}, showed that these materials exhibit first-order phase transitions at low temperatures \cite{kleemann}.

Thus, it has been claimed that \textit{nonequilibrium} models may be relevant to explain the behavior of certain materials involving microscopic disorder such as spin glasses and random-field systems (see \cite{alonso,torres} and references therein). In the case of disorder in the magnetic field, the nonequilibrium model is defined by the Hamiltonian \cite{dickman}
\begin{equation}\label{eq1}
\mathcal{H}=-J\sum_{\langle i,j\rangle}s_{i}s_{j}-\sum_{i}h_{i}s_{i}~,
\end{equation}
\noindent
where the first sumation $\sum_{\langle i,j\rangle}$ represents the (constant) exchange interaction between nearest-neighbor pairs of spins and the second term models the effects of random magnetic fields. The nonequilibrium system is described at each time by Hamiltonian (\ref{eq1}), and the random fields $h_{i}$ are spatially distributed according to a given probability distribution $P(h_{i})$. This corresponds to the well-known equilibrium RFIM, but $h_{i}$ is continuously changed by the kinetics in such a way that it always mantains itself as a realization of $P(h_{i})$. As discussed in \cite{lacomba2,alonso}, this kind of dynamics induces randomness and a sort of dynamical frustration that does not occurs in equilibrium models. In other words, one may assume that spins and fields behave independently of each other so that a conflict occurs, and a steady nonequilibrium condition prevails asymptotically. This is consistent with the reported observation of nonequilibrium effects, for example, the influence of the details of the dynamical process (kinetics) on the steady states in some real systems (see \cite{lacomba2,dickman} and references therein).

The study of nonequilibrium models defined by Eq. (\ref{eq1}) reveals many interesting features \cite{dickman}, with a rich variety of phase transitions and critical phenomena. The known above-discussed results reveal that the critical behavior is nonuniversal, but it generally depends on apparently irrelevant details of the dynamics, like diffusion of impurities, i.e., the properties of the distribution of the random variables, and the transition rates chosen. In addition, nonequilibrium models may be relevant to describe theoretically some real systems \cite{dickman}, as a magnetic material under the action of a random (or very rapidly fluctuating) magnetic field, i.e., a field that varies according to a given probability distribution with a period shorter than the mean time between successive transitions that modify the spin configuration, or a disordered system with fast and random diffusion of impurities like in random-field systems \cite{lacomba1}.

Led by these motivations, we have studied the Nonequilibrium random-field Ising model (NRFIM) on a cubic lattice with nearest-neighbors interactions and in the presence of a random magnetic field that follows a bimodal probability distribution. We performed Monte Carlo simulations on lattices with sizes up to $L=60$ and our results suggest that the phase transitions are continuous in the high-temperature and low-field region, becoming of first-order type at low temperatures and high disorder strengths. In the low-field region, a finite-size scaling (FSS) analysis shows that the system follows standard FSS laws, as was claimed in \cite{meu_jstat}.


\section{Model and Monte Carlo Simulation}

We have considered a random-field system described by the Hamiltonian (\ref{eq1}) on a cubic lattice of linear dimension $L$, with the random field $\{h\}$ following a bimodal probability distribution,
\begin{equation}\label{eq2}
P(h)=\frac{1}{2}\delta(h-h_{o})+\frac{1}{2}\delta(h+h_{o}),
\end{equation}
\begin{figure}[t]
\begin{center}
\includegraphics[width=0.4\textwidth,angle=0]{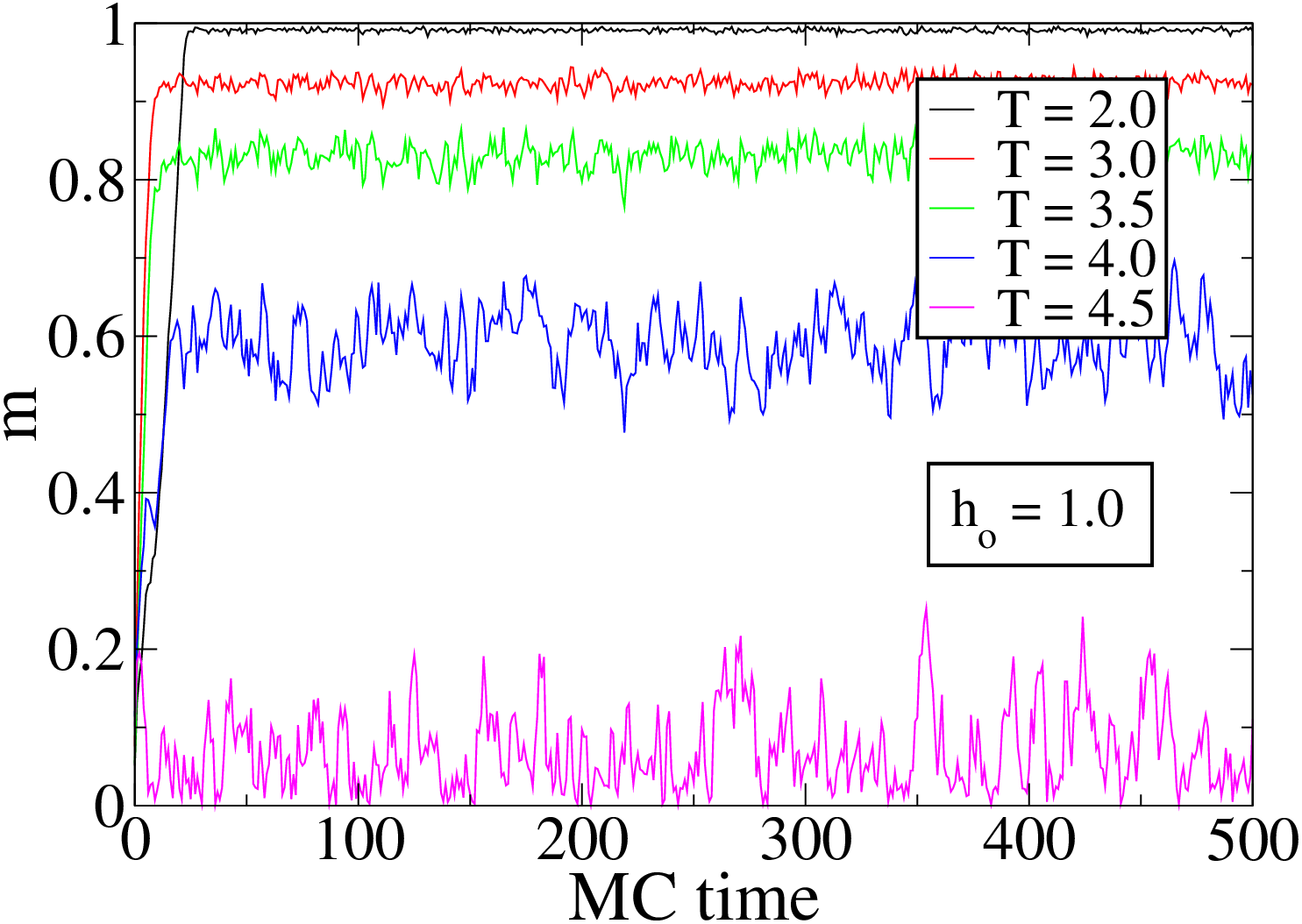}
\vspace{0.5cm}
\includegraphics[width=0.4\textwidth,angle=0]{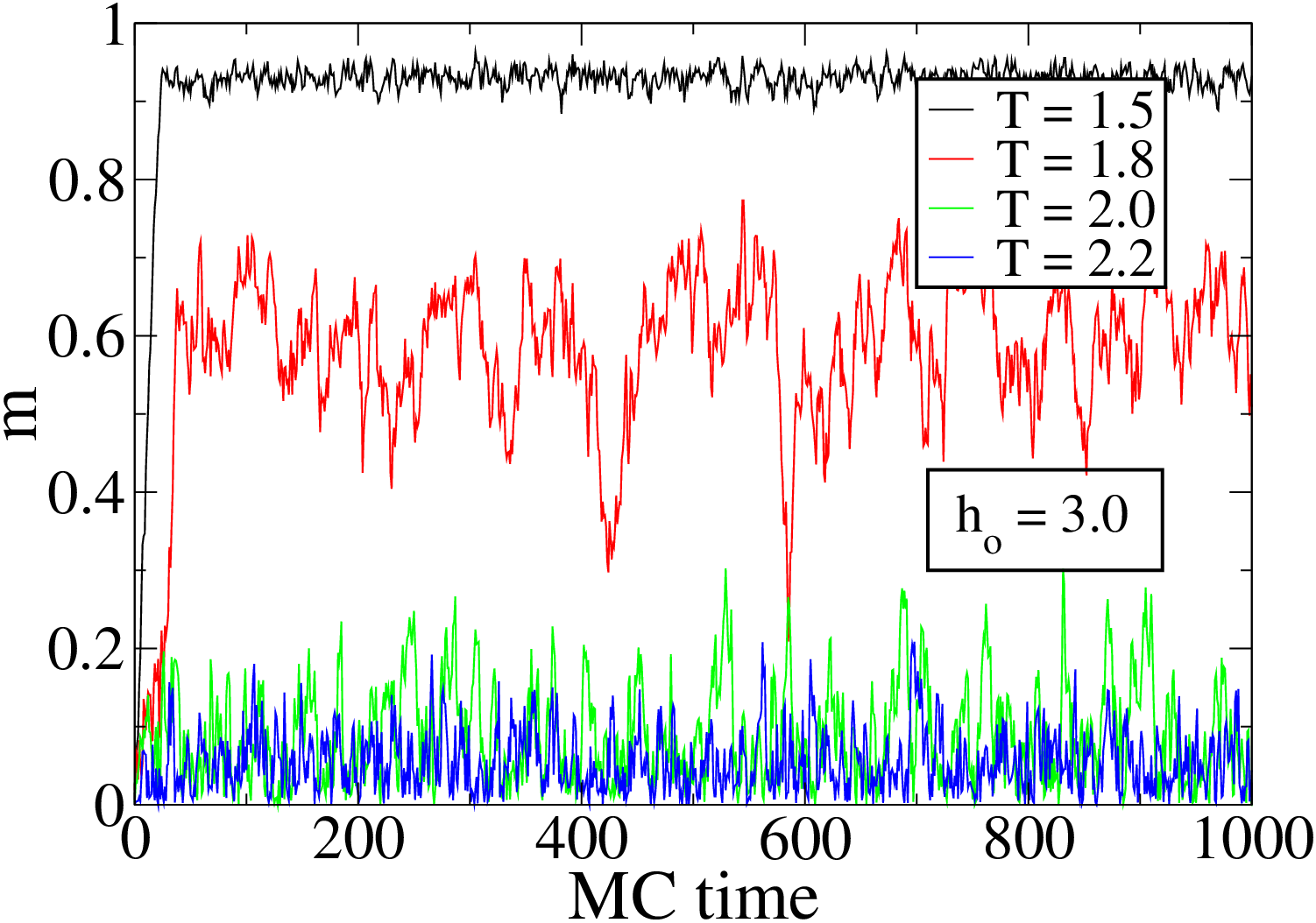}
\end{center}
\caption{(Color online) Evolution of the magnetization as a function of the Monte Carlo time, for $L=16$. Examples for low (left side) and high disorder strength $h_{o}$ (right side) are shown. As discussed in the text, the steady states are easily achieved for low and high $h_{o}$.}
\label{Fig1}
\end{figure}

\noindent
whose results in the mean-field approximation \cite{alonso} and on square lattices \cite{meu_jstat} obtained by numerical implementations constitute an interesting means of comparison for our own outcome. Any configuration $\textbf{s}=\{s_{i}\}$ evolves stochastically with time by spin flips with rate
\begin{equation}\label{eq3}
\omega(s_{i}\to -s_{i}) = {\rm min}\{1,\exp(-\delta\mathcal{H}_{i}/T)\}~,
\end{equation}
\noindent
which corresponds to the Metropolis' algorithm \cite{metropolis}, and $\delta\mathcal{H}_{i}$ stands for the flip energy cost (we set the Boltzmann constant to unity). For the numerical implementation, the algorithm is as follows: at each time step, a new configuration of random fields $\{h\}$ is generated according to $P(h)$, Eq. (\ref{eq2}); then, every lattice site is visited, and a spin flip occurs according to rate (\ref{eq3}). In other words, the random variables $h$ vary with time, i.e., the system is described at each time by Eq. (\ref{eq1}), with $h$ distributed according to $P(h)$ given by Eq. (\ref{eq2}). Thus, we have two different characteristic time scales: one for the fluctuations of the spins and another one for the fluctuations of the random field, and for simplicity we have considered that these two fluctuations are independent (a formal discussion about this is found in ref. \cite{dickman}, chapter 7).

In the following we use for simplicity $J=1$. We have studied systems of $L=10, 12, 16, 20, 24, 30, 40$ and $60$ with periodic boundary conditions and a random initial configuration of the spins. We have analyzed the following values of the parameters: $0.0<h_{o}<5.0$ and $0.01<T<10.0$. The results for all values of the magnetic field parameter $h_{o}$ show that finite-size effects are less-pronounced for $L\geq 16$. We can test the equilibration of the system by monitoring the magnetization as a function of the MC time. We have found that the steady states are easily achieved for low and high $h_{o}$, as shown in Fig. \ref{Fig1}. Thus, we have used $5\times 10^{3}$ MC steps for equilibration and $3\times 10^{6}$ MC steps for averaging.

\begin{figure}[t]
\begin{center}
\vspace{1.0cm}
\includegraphics[width=0.45\textwidth,angle=0]{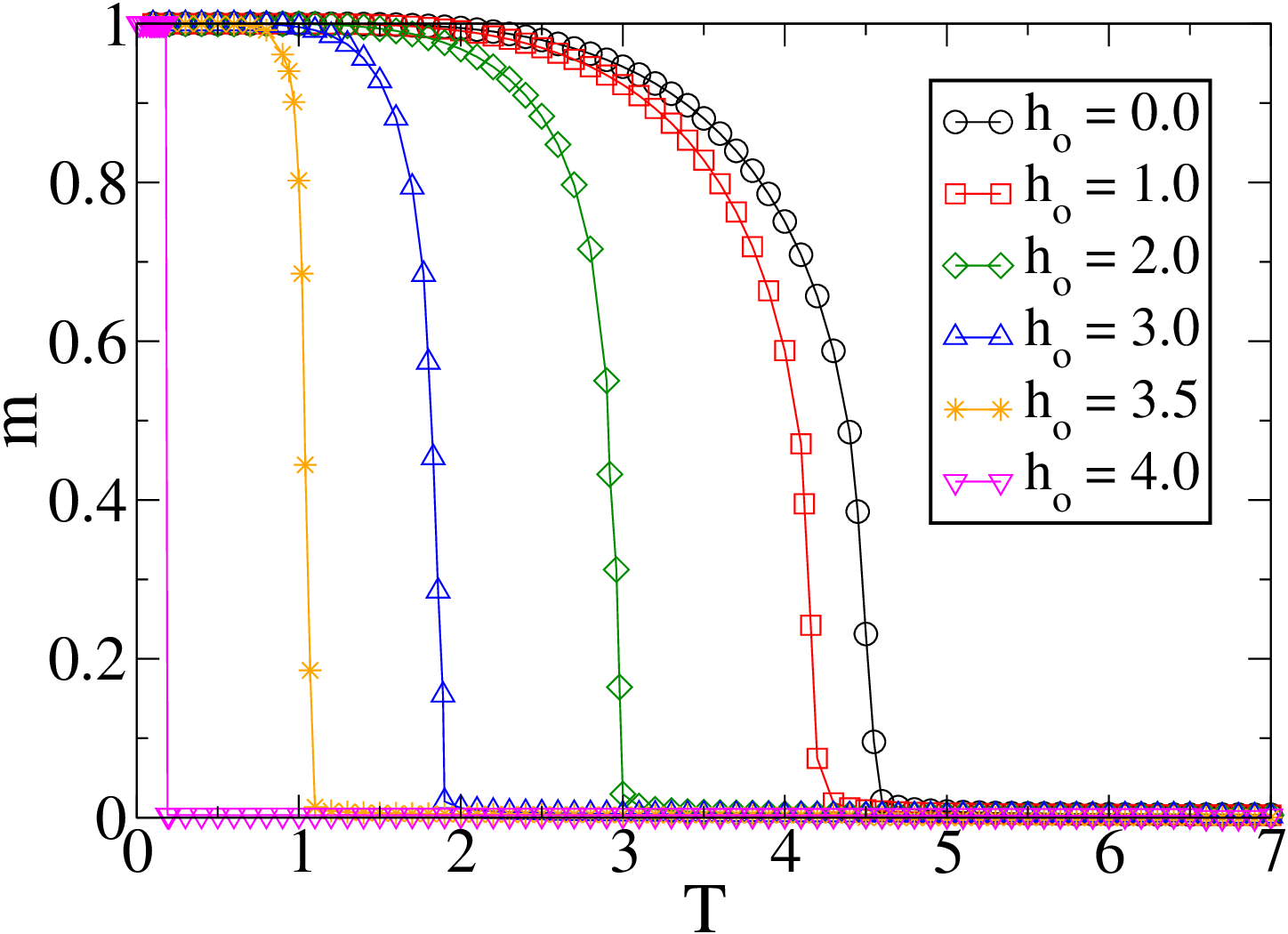}
\end{center}
\caption{(Color online) Magnetization versus temperature for $L=60$ and typical values of the disorder strength $h_{o}$. We can observe continuous phase transitions between the ordered and the disordered phases for small values of $h_{o}$, but for values near $h_{o}=4.0$ we have discontinuous transitions. For higher disorder the system is in the paramagnetic state.}
\label{Fig2}
\end{figure}

\begin{figure}[t]
\begin{center}
\includegraphics[width=0.35\textwidth,angle=0]{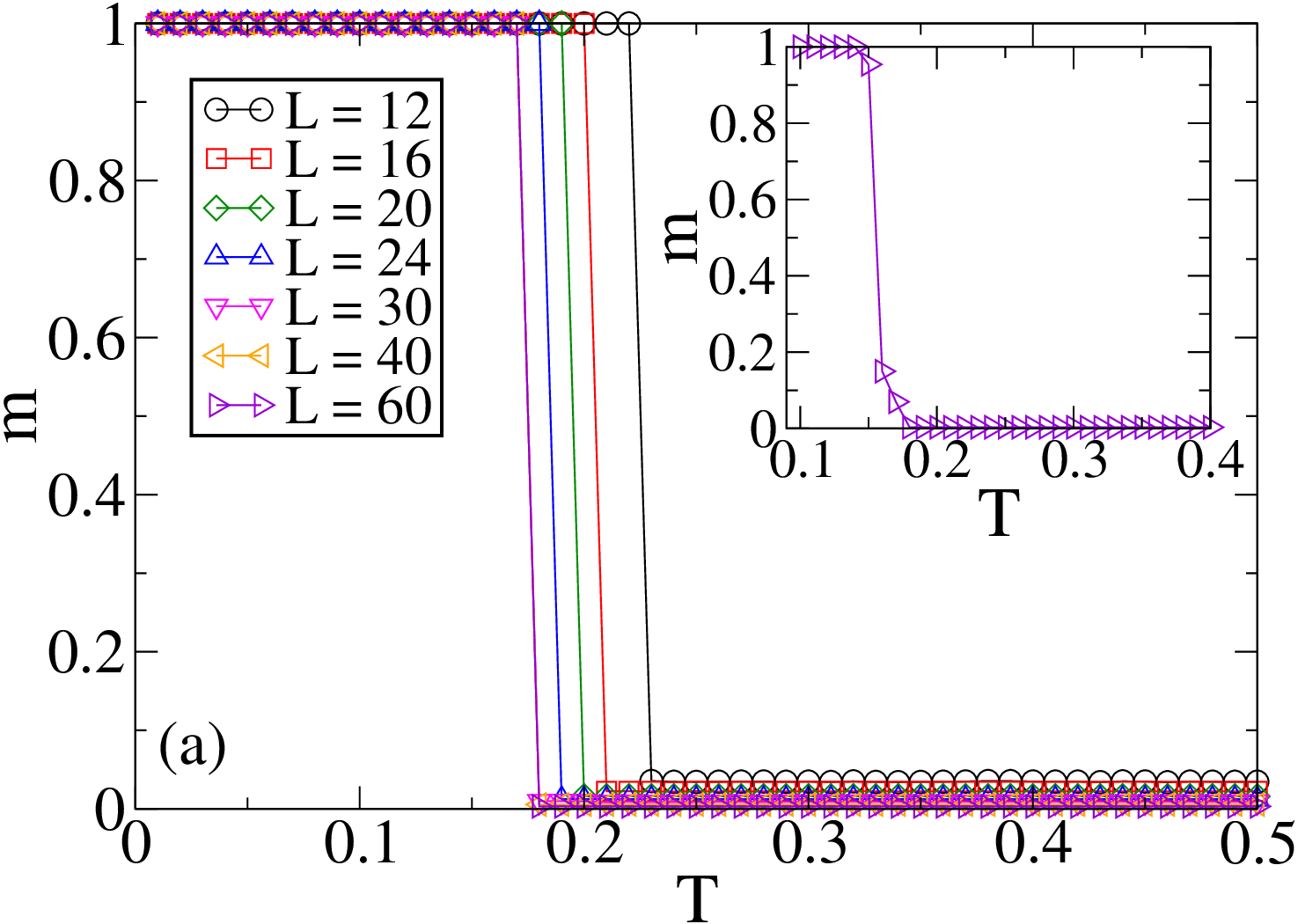}
\hspace{1.0cm}
\includegraphics[width=0.35\textwidth,angle=0]{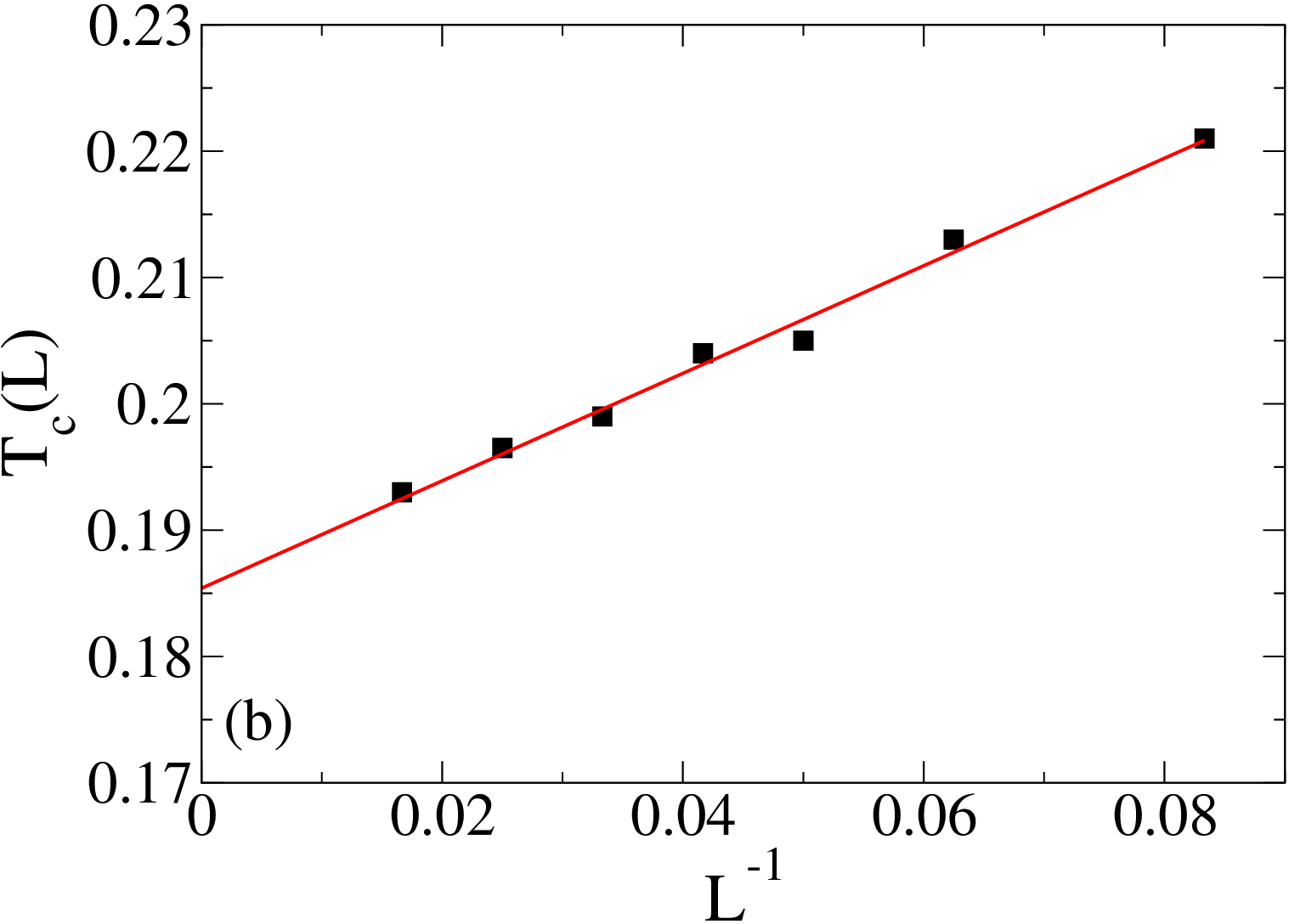}
\\
\vspace{1.0cm}
\includegraphics[width=0.35\textwidth,angle=0]{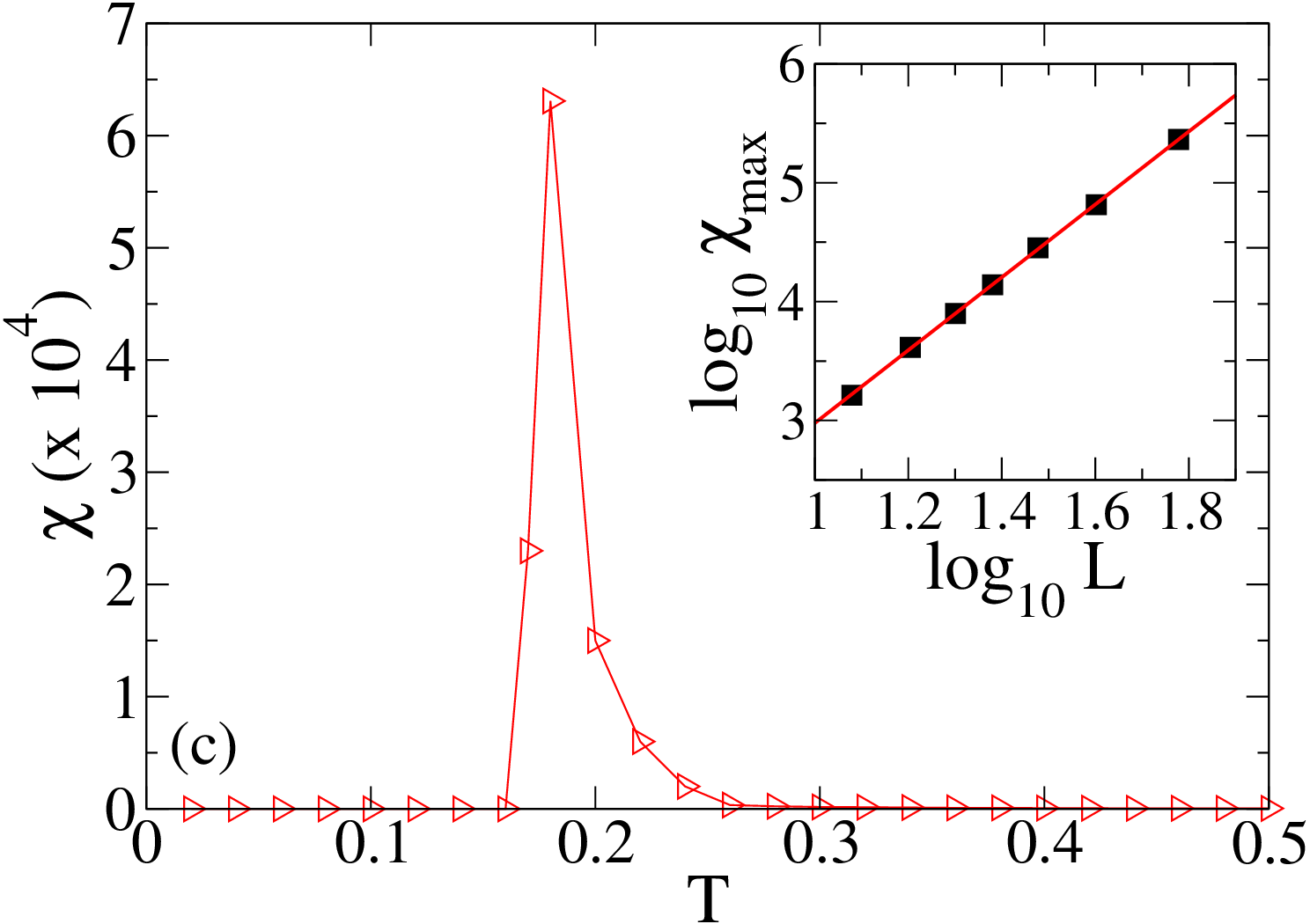}
\hspace{1.0cm}
\includegraphics[width=0.35\textwidth,angle=0]{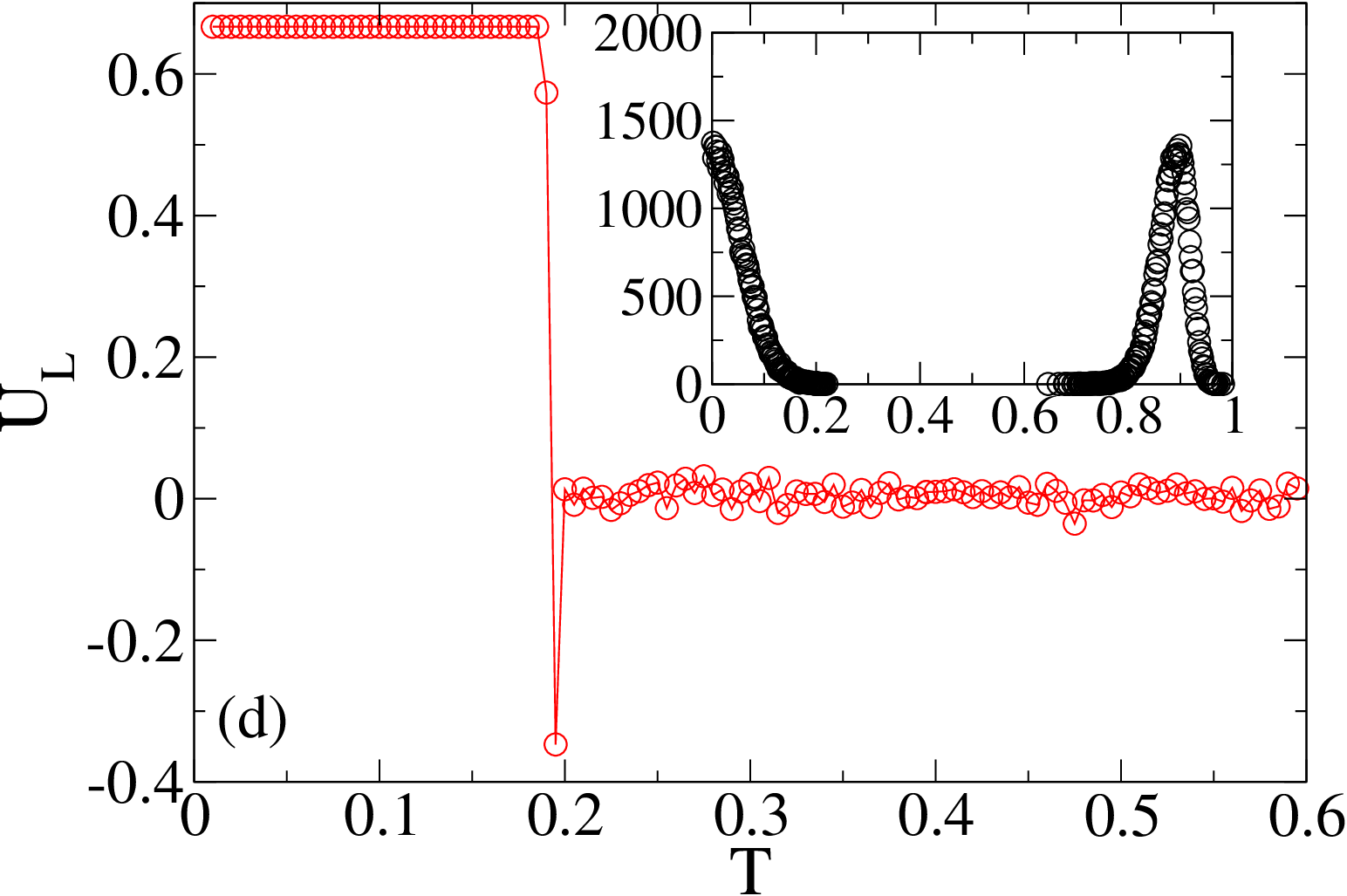}
\end{center}
\caption{(Color online) Upper figures: magnetization as a function of temperature for $h_{o}=4.0$ and for some lattice sizes $L$, showing discontinuous transitions at the pseudo-critical temperatures $T_{c}(L)$ (a). The inset shows simulations for $L=60$ using a smaller interval between temperatures, where we can see points in the coexistence region (with $0<m<1$). It is also shown an ilustration of the extrapolation procedure to determine the critical temperature $T_{c}$ for $h_{o}=4.0$ (b). Fitting data, we have obtained $T_{c}(L)=0.186+0.424\,L^{-1}$, which results in $T_{c}(L)=0.186$, in the thermodynamic limit ($L^{-1}\to 0$). Lower figures: susceptibility $\chi(T)$ as a function of temperature for $L=40$ and $h_{o}=4.0$, where it is shown a jump near the critical temperature (c). In the inset it is also shown the susceptibility peaks positions $\chi_{max}$ versus linear lattice size in the log-log scale. The slope of the straight line is $3.05\pm 0.05$, which suggests a first-order transition for $h_{o}=4.0$. This first-order character is confirmed by the behavior of the Binder cumulant, which presents a characteristic well-defined minimum near the transition (d). It is also shown in the inset the double-peaked histogram of the magnetization near the critical temperature.}
\label{Fig3}
\end{figure}

Results for the magnetization per spin as a function of the temperature are shown in Fig. \ref{Fig2}. The results suggest that the system goes continuously from the ferromagnetic ($m=1$) to the paramagnetic steady state ($m\to 0$) for values of the disorder strength $h_{o}<\sim 3.7$. For increasing values of $h_{o}$, the transition between the steady states become discontinuous, which is a indicative that the transition may be of first-order type. For $h_{o}>\sim 4.3$ the system is in the paramagnetic phase.

\begin{figure}[t]
\begin{center}
\includegraphics[width=0.4\textwidth,angle=0]{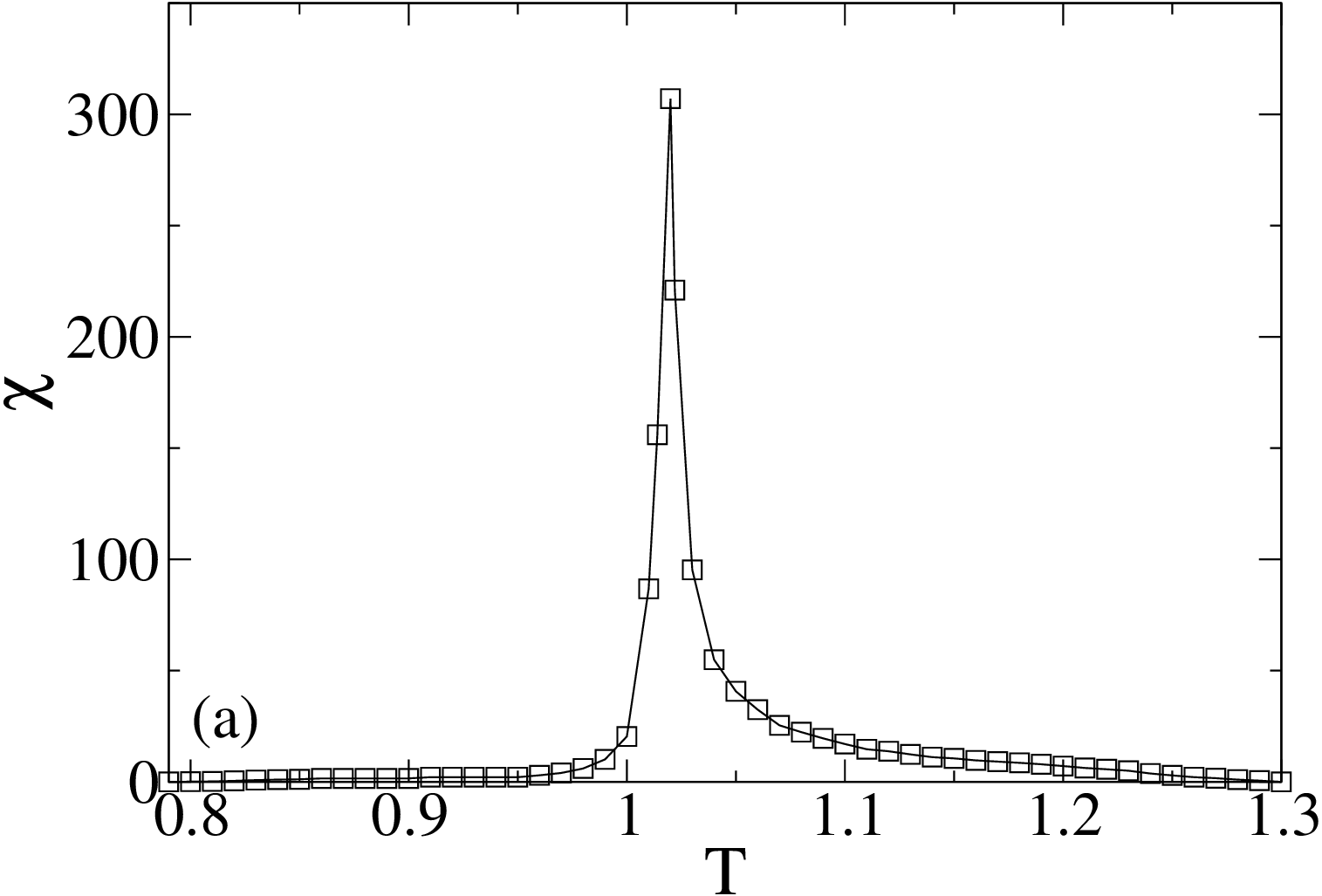}
\hspace{0.5cm}
\includegraphics[width=0.4\textwidth,angle=0]{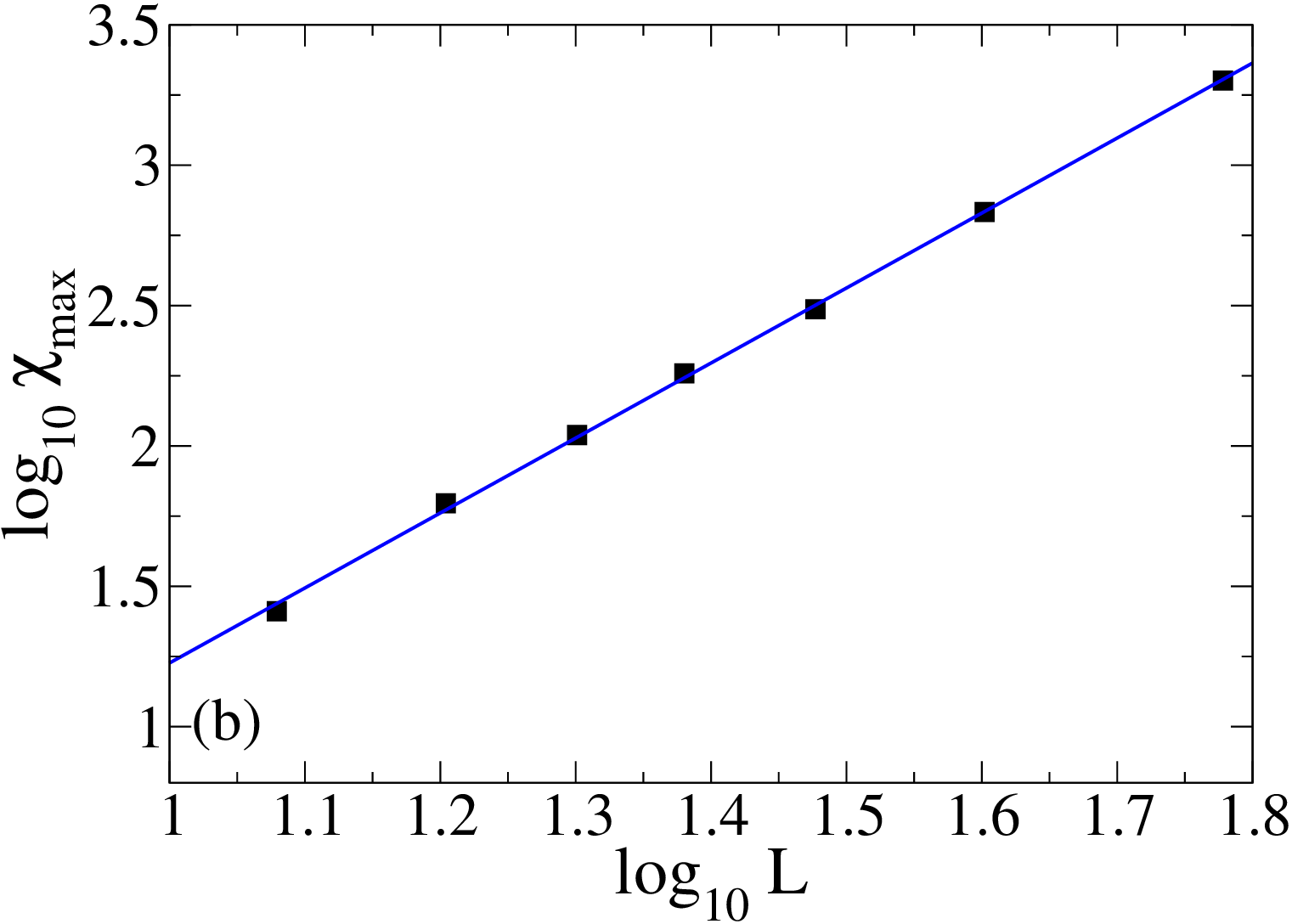}
\end{center}
\caption{(Color online) Susceptibility $\chi(T)$ as a function of temperature for $h_{o}=3.5$ and $L=40$ (a) and the peaks positions $\chi_{max}$ versus lattice size $L$ in the log-log scale (b). The slope of the straight line is $2.63\pm 0.10$, which indicates a continuous transition for $h_{o}=3.5$.}
\label{Fig_new}
\end{figure}

For the correct characterization of the discontinuous transition, we must analyze the fluctuations of the order parameter, given by
\begin{equation}\label{eq7}
\chi(T)=L^{3}\;\frac{\langle m^{2}\rangle-\langle m\rangle^{2}}{T},
\end{equation}
where $\langle\;\rangle$ stands for MC or time average \cite{miranda}. The susceptibility as a function of the temperature is shown in Figs. \ref{Fig3}(c) and \ref{Fig_new}(a), where we can easily distinguish two different behaviors, for discontinuous ($h_{o}=4.0$) and continuous transitions ($h_{o}=3.5$), respectively. As discussed in \cite{meu_jstat}, the susceptibility peaks positions grow with the system size as
\begin{equation}\label{eq8}
\chi_{max}\sim L^{a},
\end{equation}

\noindent
where $a=d$ for first-order transitions \cite{fisher_berker}, $d$ is the dimension of the lattice ($d=3$ in our case) and $a=\gamma/\nu<d$ for continuous transitions. Fitting data of $\chi_{max}$ we have found, for $h_{o}=4.0$, that [see the inset of Fig. \ref{Fig3}(c)]
\begin{equation}\label{eq9}
\chi_{max}\sim L^{b},
\end{equation}
\noindent
where
\begin{equation}\label{eq10}
b=3.05\pm 0.05,
\end{equation}

\begin{figure}[t]
\begin{center}
\includegraphics[width=0.35\textwidth,angle=0]{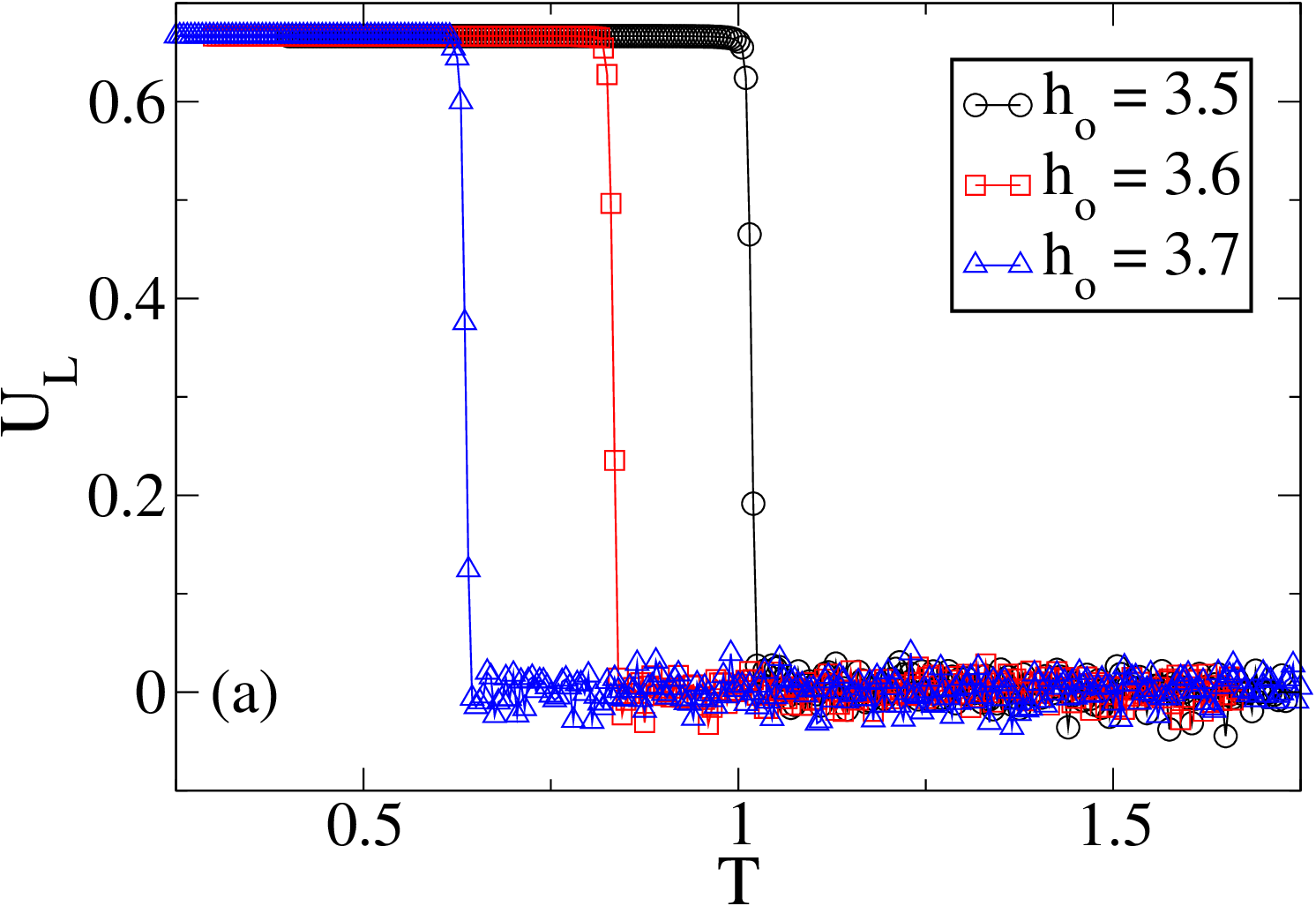}
\hspace{1.0cm}
\includegraphics[width=0.35\textwidth,angle=0]{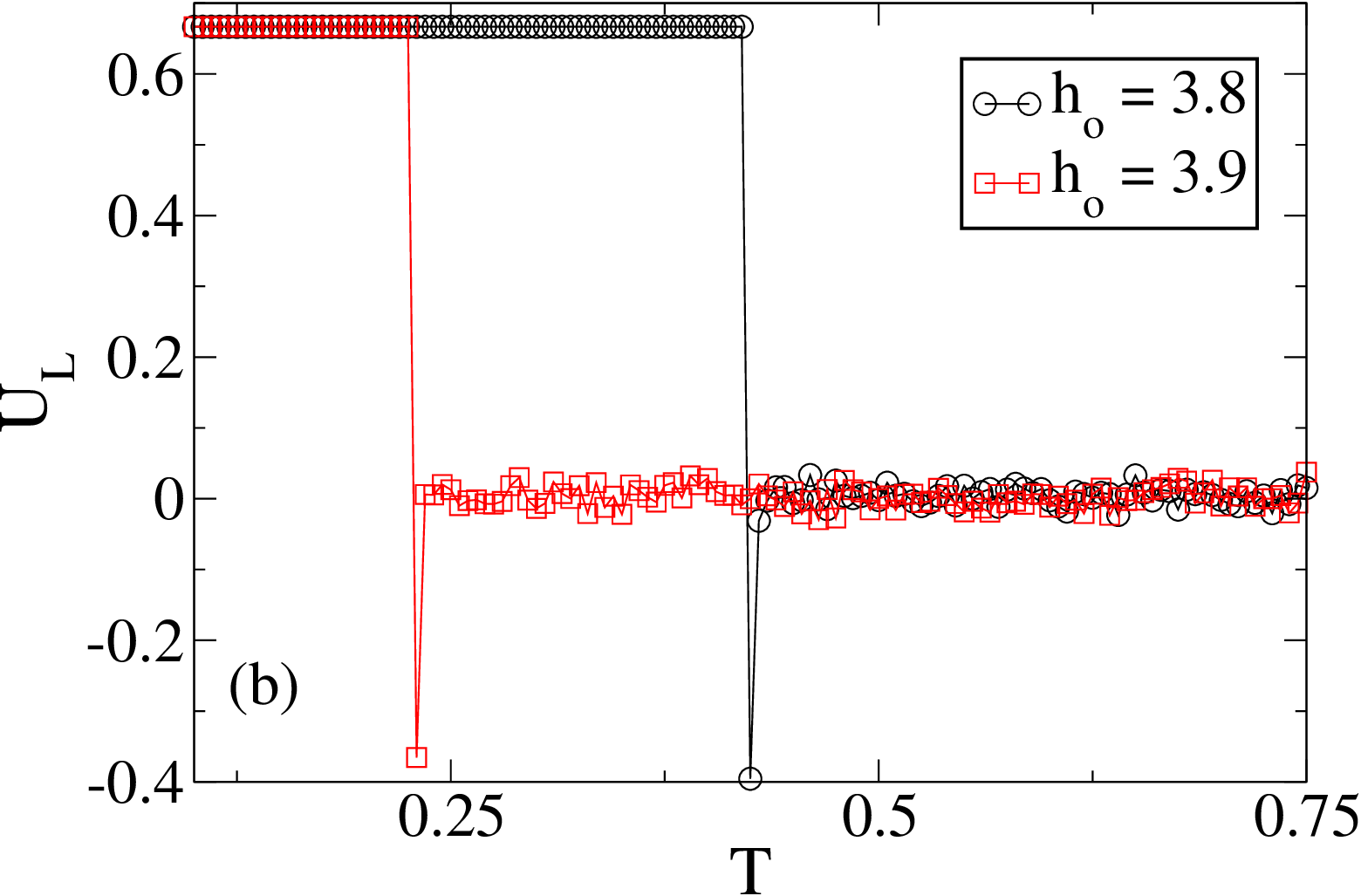}
\\
\vspace{1.0cm}
\includegraphics[width=0.35\textwidth,angle=0]{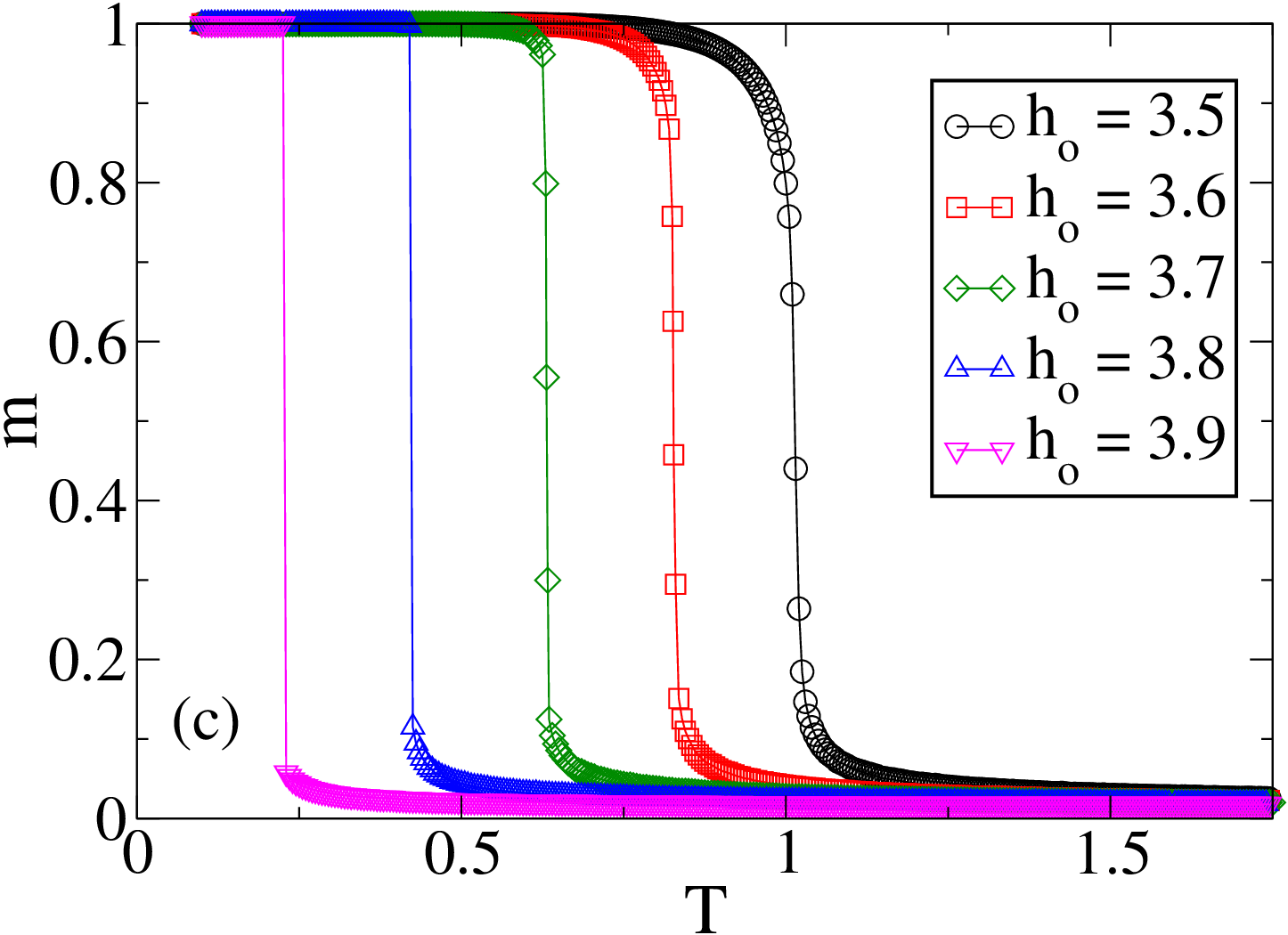}
\end{center}
\caption{(Color online) Numerical results near the tricritical point for $L=30$ and a very small interval between temperatures ($\Delta T=0.005$). In figure (a) we show the Binder cumulant for disorder strengths $h_{o}=3.5, 3.6$ and $3.7$, showing the continuous character of the transition, whereas in (b) we show the typical behavior of first-order transitions, for $h_{o}=3.8$ and $3.9$. In the lower figure we show the corresponding plots of the magnetization per spin, where we can observe jumps for $h_{o}=3.8$ and $3.9$ (c). These results suggest that the tricritical point is located in the range $3.7<h_{o}<3.8$.}
\label{FigZ}
\end{figure}

\noindent
which is compatible with a first-order phase transition. In addition, the first-order character of the transition is confirmed by the behavior of other quantities: the Binder cumulant \cite{binder}, $U_{L}=1-\frac{\langle m^{4}\rangle}{3\langle m^{2}\rangle^{2}}$, which presents a characteristic well-defined mininum near $T_{c}$ \cite{salinas} and the two-peaked histogram of the magnetization near the critical temperature [see Fig. 3(d)]. On the other hand, for $h_{o}=3.5$ we have found that $\chi_{max}\sim L^{2.63}$ [see Fig. \ref{Fig_new}(b)], confirming the continuous character of the transition [see also Fig. \ref{FigZ}(a)], as was observed earlier. The discontinuous transition also occurs in the model for other values of $h_{o}$ ($3.9, 4.1, 4.2$, for example), and the above-discussed behaviors were also observed ($\chi_{max}\sim L^{d}$, two-peaked histograms of $m$ and a well-defined minimum of the Binder cumulant), indicating that first-order phase transitions occur in the model for a small range of the disorder strength $h_{o}$, at low temperatures. On the other hand, as above-mentioned, for high temperatures the transition is continuous, indicating the occurrence of a finite-temperature nonequilibrium tricritical point, where the ordered and the disordered phases become identical. The exact location of this tricritical point is difficult to numerically determine, but the simulations suggest that this point is in the range $3.7<h_{o}<3.8$ [see Figs. \ref{FigZ}]. This scenario is quite different from the 2D \cite{meu_jstat} and the mean-field cases \cite{alonso} when the Metropolis transition rate was considered. While the 2D case is compatible with a continuous critical frontier between the ferromagnetic and the paramagnetic phases, and the mean-field approach foresees a first-order frontier, we have found for the cubic lattice a continuous transition for high temperatures and low disorder strengths and a first-order transition for low temperatures and high disorder.

\begin{figure}[t]
\begin{center}
\vspace{0.7cm}
\includegraphics[width=0.35\textwidth,angle=0]{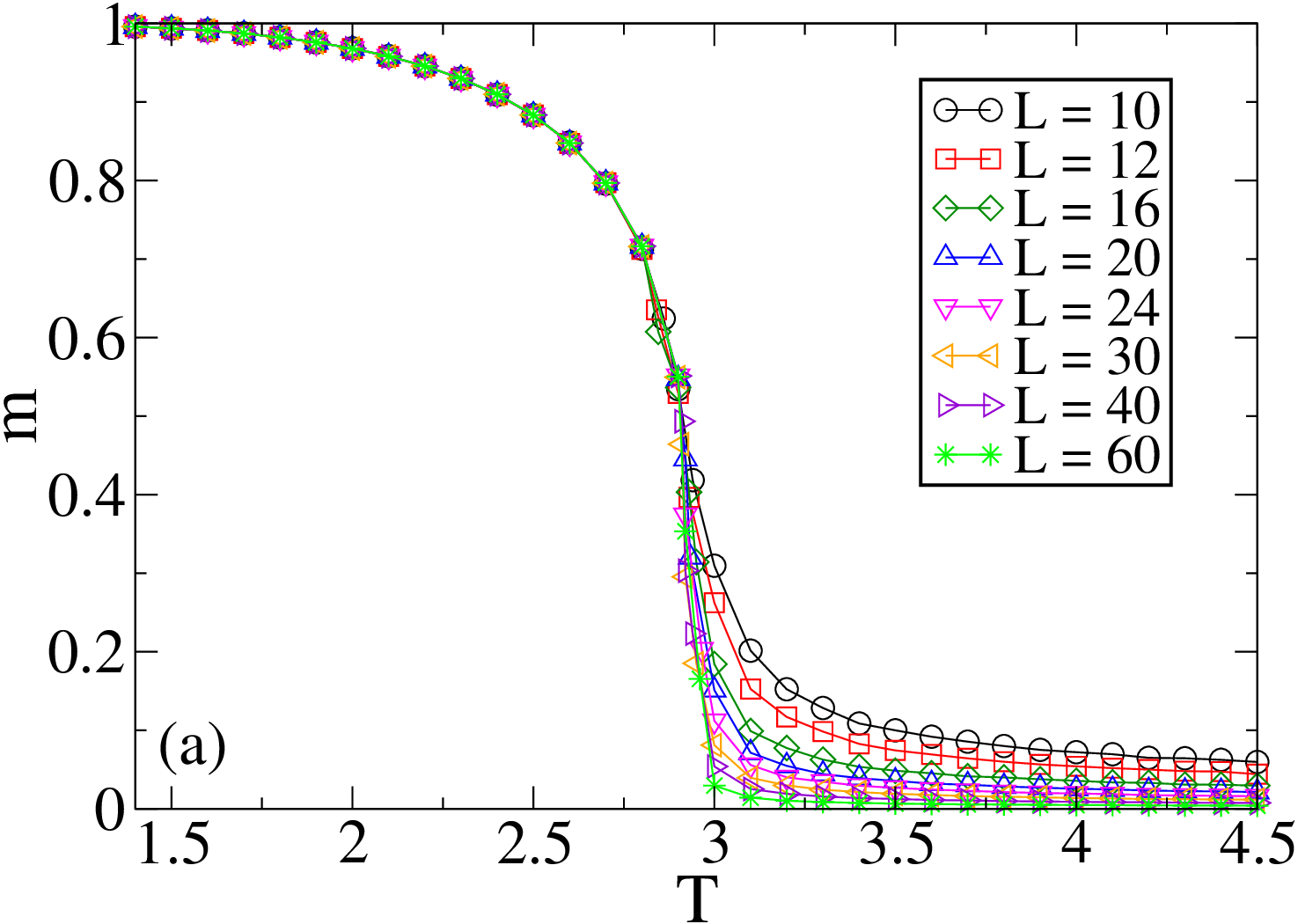}
\hspace{0.5cm}
\includegraphics[width=0.35\textwidth,angle=0]{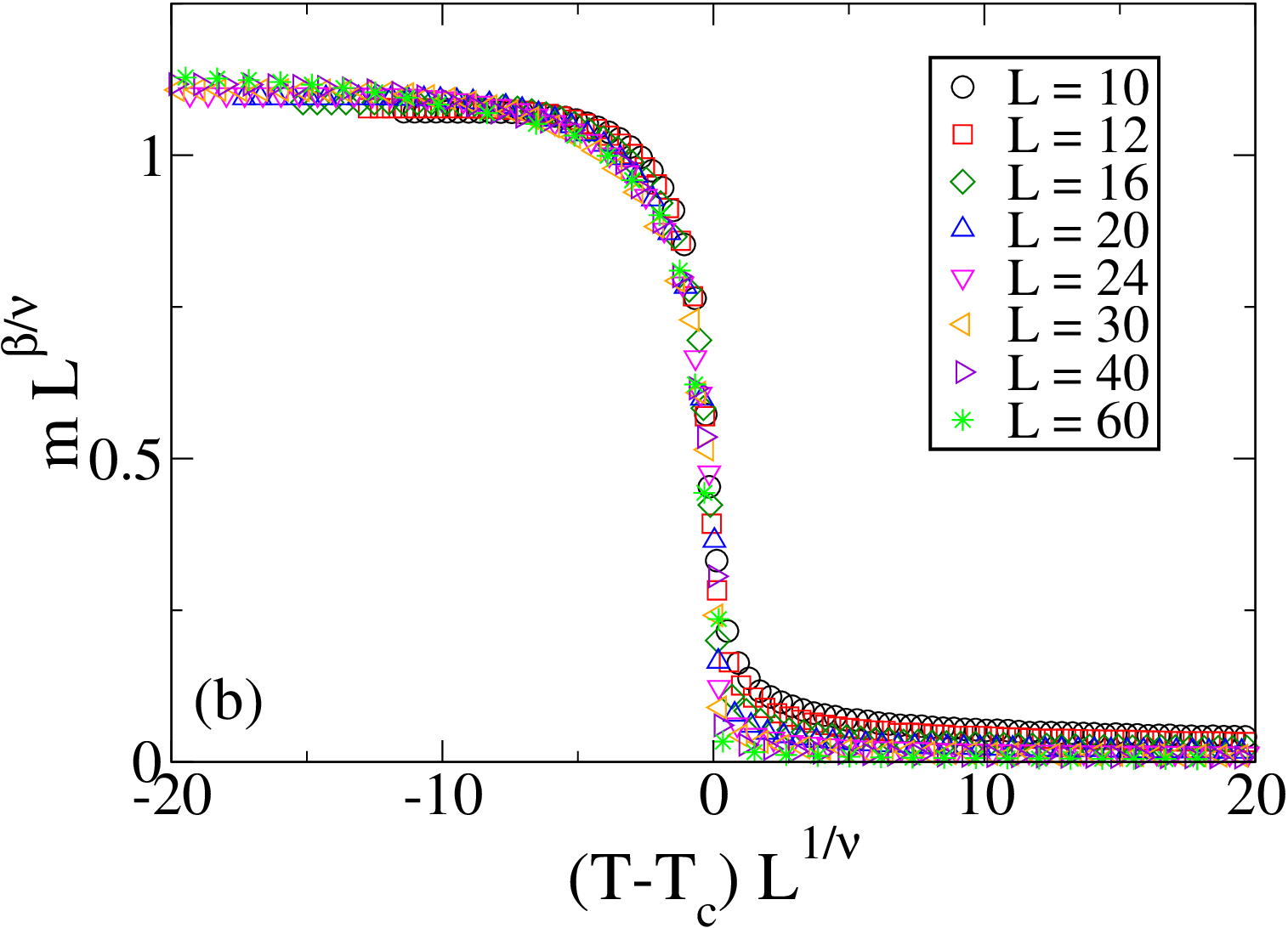}
\\
\vspace{1.0cm}
\includegraphics[width=0.35\textwidth,angle=0]{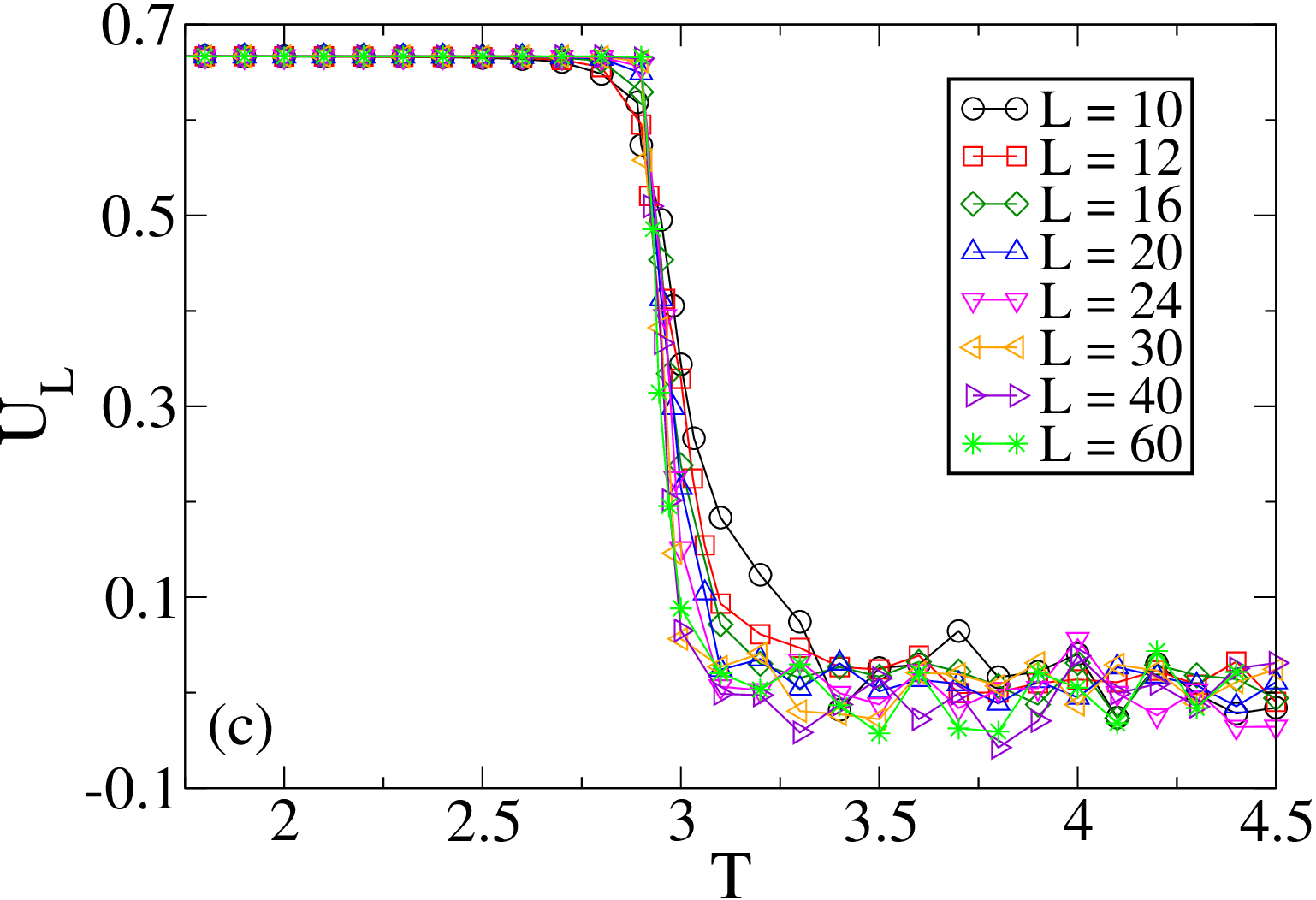}
\hspace{0.5cm}
\includegraphics[width=0.35\textwidth,angle=0]{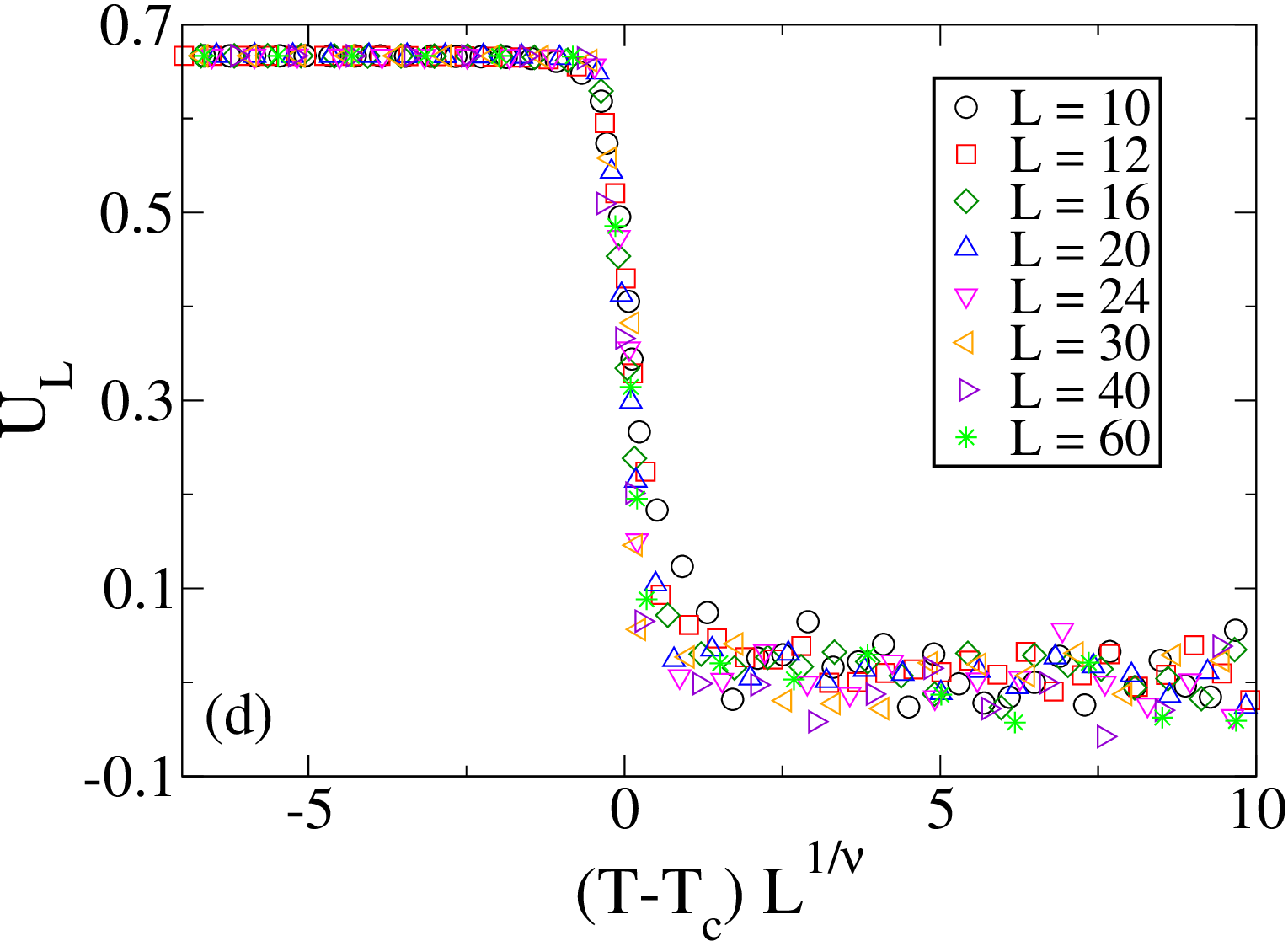}
\end{center}
\caption{(Color online) Upper figures: magnetization per spin as a function of temperature (a) and the scaling plot according to the forms in Eqs. (\ref{eq11}) for $h_{o}=2.0$ and various linear lattice sizes $L$ (b). Lower figures: Binder cumulant as a function of temperature for $h_{o}=2.0$ and some lattice sizes (c) and the best collapse of data (d), based on the finite-size scaling Eq. (\ref{eq13}). All collapses were obtained with $T_{c}=2.97$, $\beta=0.051$ and $\nu=1.67$, as discussed in the text.
}
\label{Fig4}
\end{figure}

In the 2D NRFIM \cite{meu_jstat}, the calculation of the critical exponents in the low-field region, where the phase transitions are continuous, suggests that the standard finite-size scaling (FSS) equations, i.e.,
\begin{eqnarray}\nonumber
T_{c}(L) & = & T_{c}-a\;L^{-1/\nu}, \\ \label{eq11}
m(T,L) & = & L^{-\beta/\nu}\tilde{m}((T-T_{c})\;L^{1/\nu}),
\end{eqnarray}
\noindent
are valid. Based on Eqs. (\ref{eq11}), we have calculated the critical exponents $\beta$ and $\nu$ for $h_{o}=2.0$, a disorder strength for which the simulations suggest continuous phase transitions (see Fig. \ref{Fig2}). The magnetization versus temperature curves are shown in Fig. \ref{Fig4}(a), as well as the best collapse of data [Fig. \ref{Fig4}(b)], which proves the validity of Eqs. (\ref{eq11}). The critical temperature of the infinite lattice $T_{c}=T_{c}(\infty)$ was obtained by extrapolating the $T_{c}(L)$ values given by the susceptibility peaks positions, for which we have found $T_{c}=2.97\pm 0.11$. The exponent related to the divergence of the correlation lenght $\nu$ may be calculated by means of the Binder cumulant [see Fig. \ref{Fig4}(c)], which has the FSS form
\begin{equation}\label{eq13}
U_{L}=\tilde{U_{L}}((T-T_{c})\;L^{1/\nu}),
\end{equation}
\noindent
where $\tilde{U_{L}}$ is a scaling function that is independent of $L$. We have found for this case $\nu=1.67\pm 0.03$ [Fig. \ref{Fig4}(d)]. The exponent $\beta$ was determined by the best collapse of the magnetization data [see Fig. \ref{Fig4}(b)]. Our estimate is $\beta=0.051\pm 0.003$. Thus, the results on 3D lattices confirm indications given in \cite{meu_jstat} that the nonequilibrium version of the RFIM follows standard FSS equations for continuous transitions in the low-field region.

There are important differences between the critical behavior of the NRFIM in the presence of a bimodal random field that was studied here and the corresponding quenched RFIM. Despite the debate on the existence of a first-order transition in the equilibrium RFIM still remains open, even though some recent results suggest that the first-order transition features that appear in some simulations \cite{hernandez} are due to finite-size effects \cite{fytas}, our simulations strongly suggest that a tricritical point exist in the NRFIM. In other words, the rapid fluctuation of the random field, which is peculiar of the nonequilibrium version of the RFIM, may be responsible for the occurrence of first-order transitions. If we take into account that measurements in diluted antiferromagnets like ${\rm Fe_{x}Mg_{1-x}Cl_{2}}$, which are prototypes of experimental realizations of systems under random fields \cite{belanger_review}, suggest that first-order transitions occur in these materials at low temperatures \cite{kleemann}, our results indicate that the NRFIM may be more appropriate for a theoretical description of real systems than the equilibrium RFIM.

\begin{figure}[t]
\begin{center}
\vspace{1.0cm}
\includegraphics[width=0.45\textwidth,angle=0]{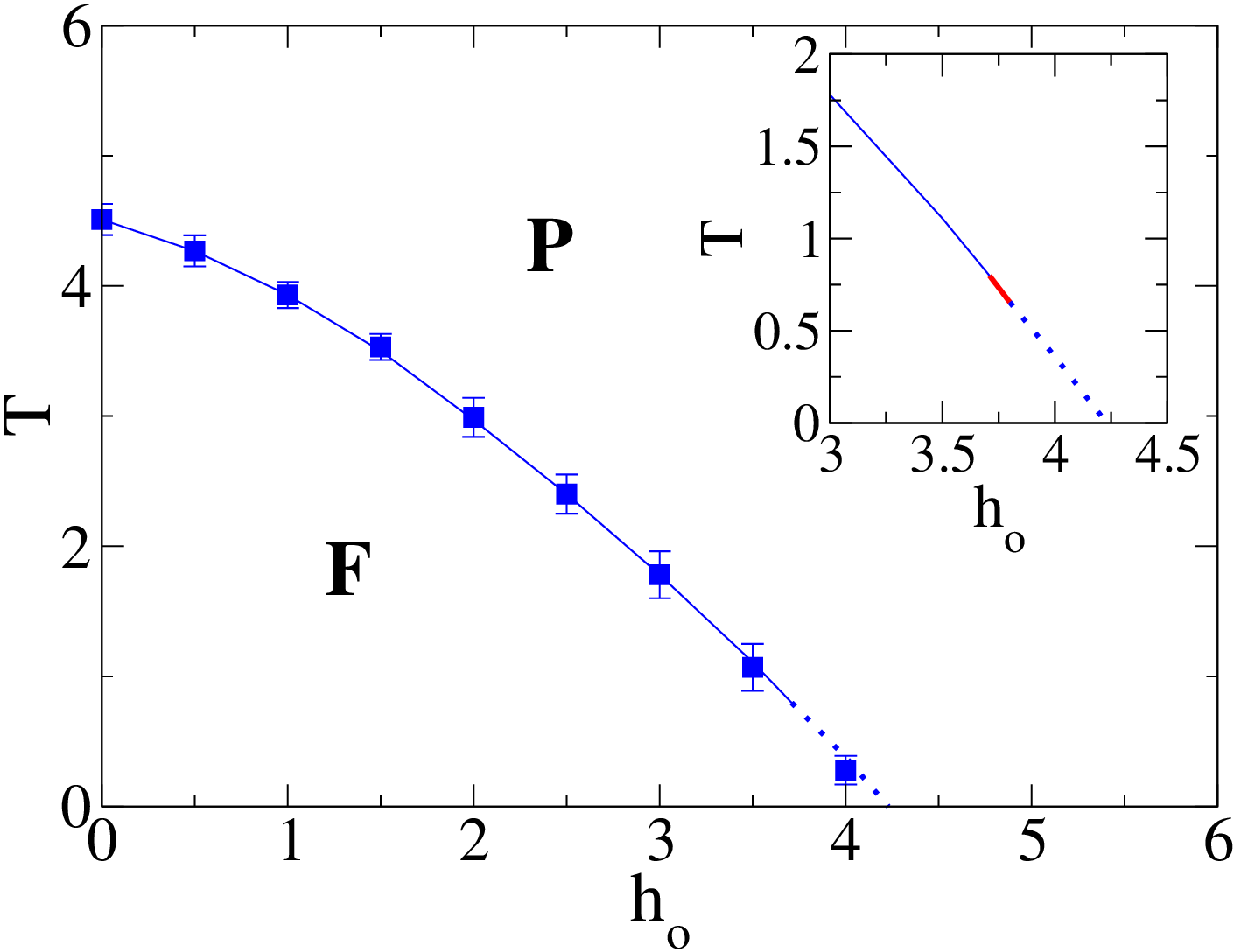}
\end{center}
\caption{(Color online) Sketch of the phase diagram of the model, separating the ferromagnetic (\textbf{F}) and the paramagnetic states (\textbf{P}). The squares are Monte Carlo estimates of the transition temperatures, whereas the line is just a guide to the eye. The dotted line at low temperatures represents the region for which the simulations suggest the occurrence of a first-order phase transition, whereas the full line represents the existence of a continuous phase transition. The inset shows in more details the region near the first-order transition (without the computed points, for clarity), and the large (red) line depicts the interval where the nonequilibrium tricritical point is expected to be located. We have obtained $T_{c}\cong 4.51$ for $h_{o}=0.0$, in agreement with the numerical estimate of the critical temperature of the 3D pure (zero-field) Ising model \cite{gould}.}
\label{Fig6}
\end{figure}

Bearing in mind the critical temperatures of the system calculated for various values of $h_{o}$, by the above-described extrapolating process, we show in Fig. \ref{Fig6} a sketch of the phase diagram of the model in the plane temperature $T$ versus disorder strength $h_{o}$, separating the ferromagnetic and the paramagnetic phases. Notice that the first-order transition that occurs at low temperatures and high disorder was not found on square lattices \cite{meu_jstat}, as previously discussed. In the same figure we show an inset where the region near the first-order transition can be seen in more details.


\section{Conclusions}

In this work, we have studied a random-field Ising model with competing kinetics defined on a cubic lattice with nearest-neighbors interactions by means of Monte Carlo simulations. The lattice sizes analyzed were $L=10, 12, 16, 20, 24, 30, 40$ and $60$. This system may be viewed as a nonequilibrium version of the random-field Ising model. We have found that the steady states are easily achieved for low and high $h_{o}$, and for averaging, we have used $3 \times 10^{6}$ MC steps.

The time evolution of the system is stochastic because of a competing spin-flip kinetics, which, in addition to the usual heat bath, involves a random external magnetic field. The competition induces a kind of dynamical frustration that might be present in real disordered systems such as the class of random-field materials \cite{lacomba2}. This system differs from the standard equilibrium ones: while the local field is randomly assigned in space according to a distribution $P(h_{i})$, which remains frozen in for the quenched case, and $P(h_{i})$ contains essential correlations in the annealed system, where the impurity distribution is in equilibrium with the spin system, our case is similar to the quenched system at each time during the stationary regime, but $h_{i}$ keeps randomly changing with time, also according to $P(h_{i})$, at each site $i$. Consequently, while frustration and randomness turn out to be rather unimportant in the annealed case, they are fundamental for the behavior of the nonequilibrium system in a way which is expected to produce macroscopic differences to the quenched case. In fact, in the annealed RFIM phase diagram \cite{claudete} the critical temperature increases as we increase the magnetic-field intensity, whereas for the equilibrium quenched RFIM \cite{fytas} and the NRFIM the disorder strength may destroy the ferromagnetic order. However, at low temperatures, it is believed that the quenched case presents continuous transitions, whereas in the nonequilibrium one our simulations suggest that first-order transitions occur.

While at mean-field level the results for the Metropolis transition rate predict a first-order phase transition in all the temperature versus disorder strength phase diagram in the case of the bimodal distribution \cite{alonso}, and MC simulations on a 2D square lattice are compatible with continuous phase transitions \cite{meu_jstat}, our numerical results suggest the occurrence of first-order transitions only at low temperatures and large disorder strengths. The order of the transition was confirmed by three quantities, that present characteristic behaviors in the case of a first-order transition: (i) the scaling of the susceptibility peaks, which grow with the total number of spins $L^{3}$ \cite{fisher_berker}; (ii) the Binder cumulant, which presents a characteristic well-defined mininum \cite{salinas}; (iii) the histograms of the magnetization, which are double-peaked \cite{salinas}. The occurrence of a first-order transition at low temperatures indicates the existence of a finite-temperature (nonequilibrium) tricritical point, whose coordinates are difficult to determine numerically though. Nonetheless, the simulations suggest that this point is located in the range $3.7<h_{o}<3.8$. The same behavior (first-order transition at low temperatures) was verified experimentally in the diluted antiferromagnets ${\rm Fe_{x}Mg_{1-x}Cl_{2}}$ \cite{kleemann}, which are prototypes of experimental realizations of systems under random fields \cite{belanger_review}.

We performed a preliminary calculation of the critical exponents in the low-field region in order to test whether the system follows continuous finite-size scaling laws, and we have found that the random-field Ising model with competing kinetics obeys standard continuous FSS equations, as suggested by Monte Carlo simulations on a square lattice \cite{meu_jstat}. As extensions of this work, it can be analyzed the case of spin-1 systems, as well as other relevant random-field probability distributions.

\vskip 2\baselineskip

{\large\bf Acknowledgments}

\vskip \baselineskip
\noindent

I am grateful to Silvio M. Duarte Queir\'os for a critical reading of the manuscript. I also acknowledge thoughtful remarks by anonymous referees which significantly improved the text. This work was supported by the Brazilian funding agency CNPq.

\end{document}